\documentclass[aps,prl,epsfigure,twocolumn,showpacs]{revtex4}
\usepackage{dcolumn}
\usepackage{bm}
\usepackage{graphicx}
\usepackage{amsmath}
\usepackage{latexsym}
\usepackage{amsfonts}
\usepackage{amssymb}
\usepackage{array}
\usepackage{times}

\newcommand{\ket}[1]{\left\vert#1\right\rangle}

\newcommand{\bra}[1]{\left\langle#1\right\vert}

\newcommand{\nbar}{\overline{n}}

\begin{document}
\title{Interaction engineering for environmental probing}
\author{M. Paternostro$^1$, S. Bose$^2$, and M. S. Kim$^1$}
\affiliation{$^1$School of Mathematics and Physics, Queen's University, Belfast BT7
  1NN, United Kingdom\\
$^2$School of Physics and Astronomy, University College London, Gower Street, London WC1E 6BT, United Kingdom}

\date{\today}

\begin{abstract}
We study the conditions for the probing of an environment affecting one party of a bipartite system of interacting objects by measurements operated only on the other element. We show that entanglement plays no crucial role in such an environment-characterization. On the other hand, if an interaction is established between the two parties, information can be reliably gathered. This result holds for both discrete and continuous variables and helps in the interpretation of recent experiments addressing the properties of mesoscopic objects.
\end{abstract}


\maketitle

``How much of the dynamics of an open system can be revealed without directly accessing it?'' This question is becoming increasingly relevant to the performance of experimental quantum information processing due to the development of a new generation of protocols dealing with systems which are only partially accessible from outside. Built-in charge qubits interacting with the field of on-chip superconducting waveguides~\cite{schoelkopf}, electron spins in quantum dots~\cite{kouwenhoven}, micro- or nano-electromechanical systems and their optomechanical counterparts~\cite{schwab,micro} and single-molecule junctions~\cite{molecule} are striking examples of the situation depicted above, where only partial access to the components of a register is possible. Another scenario where a similar question holds is given by many-body systems of (possibly interacting) particles, such as optical lattices~\cite{greiner}, where the direct addressing of a specific element of the system is difficult because of too-close inter-element distances. In this case, a winning strategy for the revelation of the dynamics of one of the particles in the register (regardless of its nature) can be the use of controllable ancillary devices acting as {detection stages}. The inability of directly addressing a system is then bypassed by coupling it to an ancilla (not necessarily being a qubit) that is easier to manipulate and detect. 

Our question has natural pragmatic relevance and is certainly experimentally stimulating. We show that it is also dressed of fundamental importance, as it can be related to the role played by entanglement in the revelation of the dynamics of a system. By studying discrete and continuous variables (CV's), we show that the engineering of {``interactions''} between the detecting stage and the system to study, rather than entanglement, is crucial in an environment-revelation process. This holds under general conditions: Entanglement alone is not sufficient for indirect environment revelation. Even in the presence of entanglement between (non-interacting) system and detecting stage, tracing the system before or after its interaction with the environment is irrelevant, when looking at the detecting part alone.
In our analysis we use a minimum-control approach, where no fast/precise interaction-switching is required and the environmental inference is performed simply by looking at the equilibrium state of the system, which is in general easily measured. These features allow for an original reinterpretation of recent experiments performed on mesoscopic optomechanical devices~\cite{micro}.

\begin{figure}[b]
\centerline{\includegraphics[width=0.18\textwidth]{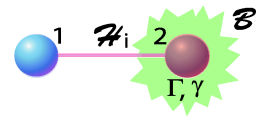}}
\caption{Sketch of the physical situation. System $1$ is completely accessible while non-accessible system $2$ is embedded in an environment, ${\cal B}$, whose parameters $\Gamma,\,\gamma$ are unknown. Systems $1$ and $2$ can interact via a non-switchable coupling ${\cal H}_i$.}
\label{fig:fig0}
\end{figure}

{\it Qubit case}.- In order to tackle the central question of our study, we start considering the simple case of two qubits, labelled $1$ and $2$ and characterized by the respective transition frequency $\omega_{j}$ ($j=1,2$). The free evolution of the qubits is therefore ruled by the Hamiltonian $\hat{\cal H}_{f}={\omega_{1}}\hat\sigma_{z,1}/2+{\omega_{2}}\hat\sigma_{z,2}/2$ (we assume $\hbar=1$) with the $z$-Pauli matrices $\hat{\sigma}_{z,j}=\ket{1}_{j}\!\bra{1}-\ket{0}_{j}\!\bra{0}$ and $\{\ket{0},\ket{1}\}_j$ the computational states of qubit $j$. Qubit $2$ embodies the system whose dynamics we want to characterize through the detecting stage represented by qubit $1$. In order to account for the case where qubit $2$ is originally part of a multipartite register whose state we are not interested in, we allow for $2$ to be prepared in a general mixed state. This would be the case, for instance, when this qubit is originally entangled with the rest of a register.
The second assumption we make is that qubit $1$ is in a pure state, well isolated from the environment affecting $2$. Detection-windows can be opened on $1$, during the dynamical evolution. This assumption does not affects the generality of our results and can be easily relaxed.

We consider a non-switchable interaction between $1$ and $2$ described by the coupling Hamiltonian $\hat{\cal H}_i$. As qubit $2$ and its surrounding physical system are unavailable to direct investigation, we assume no knowledge about the environment ${\cal B}$ affecting qubit $2$. The only working condition we take is that ${\cal B}$ is Markovian. This simplification, allowing us to treat the problem with a Liouvillian approach, holds in many physical scenarios such as high-temperature circuit-QED, where the Bloch-Redfield formalism must be pushed beyond the standard secular approximation and ``quantum optics'' master equations (ME's) become the most appropriate tool~\cite{rau}. When this assumption cannot be made (as for the case of long-memory nuclear spin baths in semiconductor quantum dots), existing tools such as effective Hamiltonian approaches or non-Markovian ME could be used~\cite{myungnonmark}.

The choice of $\hat{\cal H}_i$ is a setup-dependent issue. Its form is usually guided by naturally (or easily) realized interactions specific of a chosen implementation. However, our scheme is explicitly {hybrid} and thought as the result of the combination of register and detection stage having different physical nature. As such, it is reasonable to assume the ability of engineering the most suitable form of coupling, possibly by changing the type of detection stage being used. To illustrate the main point of our analysis, we find instrumental the study of two particular models. The first is an anisotropic $xy$ coupling, which has been largely used in problems of quantum dynamics in spin chains. We take it because it naturally arises in quantum electrodynamics-like systems where two qubits dispersively interact with a bus via the off-resonant Jaynes-Cummings model~\cite{imamoglu}. This latter is almost ubiquitous where a spin-boson interaction is arranged~\cite{wilson-rae}. The $xy$ model reads $\hat{\cal H}_{xy}=J_{x}\hat{\sigma}_{x,1}\hat{\sigma}_{x,2}+J_{y}\hat{\sigma}_{y,1}\hat{\sigma}_{y,2}$. The second model we consider, $\hat{\cal H}_{{cbf}}=({g}/{2})({\openone}_{2\times{2}}+\hat{\sigma}_{z,1})\hat{\sigma}_{x,2}$, describes a bit-flip of qubit $2$ conditioned on the state of qubit $1$. Its use is related to the extension to CV's that we perform in the last part of this paper. Here, $J_{x,y}$ and $g$ denote coupling strengths. The non-unitary part of the evolution is entirely ascribed to qubit $2$ being exposed to its environment. We consider a general case of ${\cal B}$ being both dissipative and dephasing. The Liouvillian $\hat{\cal L}_{B}(\varrho)$ acting on the density matrix $\varrho$ of the system is therefore~\cite{wallsmilburn}
$\hat{\cal L}_{B}(\varrho_{})=-\Gamma(\nbar+1)(\{\ket{1}_{2}\!\bra{1},\varrho\}-2\ket{0}_{2}\!\bra{1}\varrho\ket{1}_{2}\!\bra{0})-\Gamma\nbar(\{\ket{0}_{2}\!\bra{0},\varrho\}-2\ket{1}_{2}\!\bra{0}\varrho\ket{0}_{2}\bra{1})-\gamma[\hat{\sigma}_{z,2},\hat{\sigma}_{z,2}\varrho]$.
The first two terms account for a dissipation occurring at a rate $\Gamma$ ($\nbar$ is the average thermal occupation number of the environment), proportional to its temperature, while the last term describes dephasing at rate $\gamma$. With this notation, the evolution of the density matrix of the system is given by the ME $\partial_{\tau}\varrho=-i[\hat{\cal H}_{t},\varrho]+{\cal L}_{\cal B}(\varrho)$
with $\hat{\cal H}_t=\hat{\cal H}_f+\hat{\cal H}_i$ and $\hat{\cal H}_{i}=\hat{\cal H}_{xy}$ ($\hat{\cal H}_{i}=\hat{\cal H}_{cbf}$) for the choice of the first (second) model. $\hat{\cal H}_{xy}$ can be written as $\hat{\cal H}_{xy}=\hat{\sigma}_{+,1}(J\hat{\sigma}_{-,2}+\delta\hat{\sigma}_{+,2})+h.c.$, where $J=J_{x}+J_{y}$, $\delta=J_{x}-J_{y}$ and $\hat{\sigma}_{\pm,j}=(\hat{\sigma}_{x,j}\pm{i}\hat{\sigma}_{y,j})/2$. For $\delta\neq{0}$, $[\hat{\cal H}_{xy},\sum_{j}\hat{\sigma}_{z,j}]\neq{0}$ that prevents the total number of excitation to be a good quantum number. If we consider the $cbf$ model, instead, we have $[\hat{\cal H}_{cbf},\hat{\sigma}_{z,1}]=0$, implying that the Hilbert space is partitioned in two sectors, each spanned by states having qubit $1$ in its ground or excites state, respectively. The eigenstates of $\hat{\cal H}_{cbf}$ are $\ket{00}_{12}$ and $\ket{01}_{12}$ for the first sector (eigenvalues $-\Omega/2$ and $-\bar\omega/2$) and $[-(\omega_2/2g)\mp\sqrt{1+(\omega_2/2g)^2}]\ket{10}+\ket{11}$ (with eigenvalues $(\omega_1\mp\omega_2\sqrt{1+4g^2/\omega^2_2})/2$) for the second. Here, $\Omega=\omega_1+\omega_2$ and $\bar{\omega}=\omega_1-\omega_2$. Obviously, to get entanglement through $\hat{\cal H}_{cbf}$, we should start with a superposition $\cos\vartheta\ket{0}_1+{\rm e}^{i\varphi}\sin\vartheta\ket{1}_1$ ($\vartheta\in[0,\pi/2],\,\varphi\in[0,\pi]$). 

Independently of the interaction being considered, the dynamics of the system is best evaluated by projecting the ME onto the elements of the two-qubit computational basis, {\it i.e.} by taking
$\partial_t\varrho_{lm,pq}=-i\bra{lm}[\hat{\cal H}_t,\varrho]\ket{pq}+\bra{lm}\hat{\cal L}_B(\varrho)\ket{pq}$,
where $\varrho_{lm,pq}=\bra{lm}\rho\ket{pq}$ ($l,m,p,q=0,1$) is an element of the density matrix of the two-qubit system. The explicit form of this set of coupled differential equations is straightforwardly found by direct calculation. 

By introducing the vector of density matrix elements ${\bf v}_i$ ($i=xy,\,cbf$) such that $({\bf v}_i)_{p}=\varrho_{\tilde{p}}$ with $p=0,..,15$ and $\tilde{p}$ the same number expressed in binary notation, it is possible to write such Bloch-like equations in the vectorial form $\partial_t{\mathbf v}_i={\mathbf \Sigma}_{i}{\bf v}_i$, where ${\mathbf \Sigma}_{i}$ is a $16\times16$ matrix of coefficients that is easily extracted from the Bloch equations. The dynamical evolution can be cast into the form 
${\bf v}_{i}(t)={\rm e}^{{\mathbf \Sigma}_it}{\bf v}_i(0)$, 
where the exponential matrix includes the effects of ${\cal B}$. It is straightforward to see that, in general, the steady state of the system, independently of the coupling model being assumed, can be determined as the eigenvector ${\bf v}^{ss}_i$ corresponding to the null eigenvalue of ${\mathbf \Sigma}_i$. 
For the $xy$ coupling, this immediately leads to $\varrho^{ss}_{xy}=(\Gamma/d_{xy})\tilde\varrho^{ss}_{xy}+({J^2\delta^2GN}/{d_{xy}})\openone_{4\times4}$. Here
\begin{equation}
\label{stationaryXY}
\tilde\varrho^{ss}_{xy}=
\begin{pmatrix}
\frac{(N+1){\cal A_-}}{4}&0&0&N{\cal F}\\
0&\frac{(N-1){\cal A_-}}{4}&N{\cal E}&0\\
0&N{\cal E}^*&\frac{(N+1){\cal A_+}}{4}&0\\
N{\cal F}^*&0&0&\frac{(N-1){\cal{A_+}}}{4}
\end{pmatrix}
\end{equation}
with $N=2\nbar+1,\,G=\Gamma{N}+2\gamma$ and, explicitly, ${\cal A}_\pm=(N\mp{1})(G^2+{\bar\omega}^2)\delta^2+(N\pm1)(G^2+\Omega^2)J^2$,  ${\cal E}=-i\Gamma{JN}\delta^2(G+i\bar\omega)$, ${\cal F}=i\Gamma{J}^2\delta(G-i\Omega)$ and $d_{xy}=N\{\Gamma{N}[(G^2+\omega^2)\delta^2+(G^2+\Omega^2)J^2]+4GJ^2\delta^2\}$. The steady state $\varrho^{ss}_{xy}$ is thus a mixture of Bell states, its separability resulting from the trade-off between the parameters entering such a mixture. On the other hand, $\varrho^{ss}_{cbf}={\mathbb A}\cos^2\vartheta\oplus{\mathbb B}\sin^2\vartheta$ with ${\mathbb A}={\rm Diag}[(1+\nbar)/N,\nbar/N]$ and $d_{cbf}{\mathbb B}=\{[J^2G+\nbar\Gamma(G^2+\omega^2_2)]\openone_{2\times2}-J\Gamma(\omega_2\hat\sigma_{x,2}+G\hat{\sigma}_{y,2})+\Gamma(G^2+\omega^2_2)\ket{0}_2\!\bra{0}\}$ with $d^{-1}_{cbf}={\rm Tr}{\mathbb B}$.
It is easy to see that $\varrho^{ss}_{cbf}$ is separable. A way to understand it is the following: If we choose a state having $\vartheta=0$, for instance, we know that the state has to remain separable throughout the entire dynamics leading to the steady state. However, the properties of the steady state do not depend, by definition, on the specific instance of initial state being considered. Therefore, the only possibility we have is that the steady state is always separable. Of course, dynamically, there can be a transient during which quantum correlations are present. This is a point of interest of the following analysis.

\begin{figure}[t]
\centerline{\includegraphics[width=0.5\textwidth]{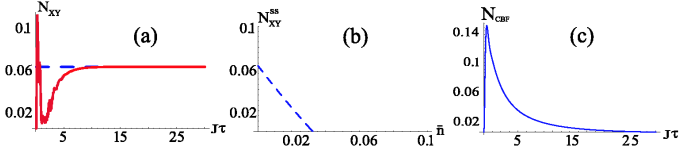}}
\caption{{\bf (a)}: $N_{xy}$ against $J\tau$ (solid line) for $J=0.3,\,\delta=0.1,\,\bar\omega=0.3,\,\Omega=3$, $\gamma=0.001,\,\Gamma=0.1$ and $\nbar=10^{-3}$. Qubit $1$ is prepared in $\ket{0}_1$. We took $100$ initial density matrices for qubit $2$, each corresponding to a Bloch vector extracted from a Gaussian sample. We plot the average entanglement of such an ensemble (solid line). The dashed line shows $N^{ss}_{xy}$. {\bf (b)}: $N^{ss}_{xy}$ against $\nbar$ for the same parameters in {\bf (a)}. {\bf (c)}: $N_{cbf}$ for $g=0.3,\,\bar\omega=0.3,\gamma=10^{-2},\,\Gamma=1,\,\Omega=3$ and $\nbar=10^{-3}$. Qubit $1$ is prepared in $(\ket{0}_1+\ket{1}_1)/\sqrt{2}$.}
\label{fig:dinamico1}
\end{figure}

The entanglement measure that we adopt here is based on negativity of the partial transposition criterion, which is necessary and sufficient in the general two-qubit case~\cite{myungDNS}. The dynamical entanglement $N_{i}(t)$, calculated via the knowledge of ${\bf v}_{i}(t)$, can be compared with entanglement at the steady state $N^{ss}_{i}$. Figs.~\ref{fig:dinamico1} show some significant examples of such a comparison. In panel {\bf (a)}, relative to the $xy$ model, $N_{xy}(t)$ is persistent in time, stabilizing at about $0.06$. This is in contrast with the situation of panel {\bf (c)}, where $N_{cbf}(t)$ (for qubit $1$ prepared in $(\ket{0}_1+\ket{1}_1)/\sqrt{2}$) goes to zero. Obviously, in a ``warm'' environment having $\nbar=1$, $N_{xy}(t)$ is no more persistent in time. This is better shown in Fig.~\ref{fig:dinamico1} {\bf (b)}, where $N^{ss}_{xy}$ is plotted against $\nbar$ and shown to be almost linearly decreasing (see the figure caption for more details). As anticipated, $N^{ss}_{cbf}$ is always zero.
For the $cbf$ Hamiltonian, in Fig.~\ref{fig:dinamico1} {\bf (c)} we show an example addressing a strong environment. An emerging feature is that the time at which the system reaches its steady state depends very weakly on the initial state of qubit $2$ but critically on the parameters of its environment. The larger $(\gamma,\Gamma,\nbar)$ with respect to $(J,\delta)$ or $g$, the quicker $\rho^{ss}_{i}$ is set, this being entangled depending on the trade-off between coherent and incoherent dynamics. 

{\it Environment probing}.- Motivated by recent efforts produced, for instance, in circuit-QED~\cite{schoelkopf}, in order to discern the effect of the coupling between qubit $2$ and its environment, we consider the spectrum of emission from qubit $1$. The choice of this figure of merit, which is standard in quantum optics, is strengthened by the fact that ``easy'' single-qubit measurements are required to reconstruct it. We aim at determining the spectrum 
$S(\nu)=\Re\left[\int^{\infty}_{0}d\tau{\rm e}^{i\nu{\tau}}\langle\hat{\sigma}_{+,1}(\tau)\hat{\sigma}_{-,1}(0)\rangle_{ss}\right]$,
where $\langle\hat{\sigma}_{+,1}(\tau)\hat{\sigma}_{-,1}(0)\rangle_{ss}$ is a {two-time correlation function} evaluated in the steady state. 
The linearity of the Bloch-like equations allows the use of the quantum regression theorem  to obtain the two-time correlation function we need~\cite{wallsmilburn}. We indicate with ${\bf C}_i(\tau)$ the vector of two-time correlations relative to a choice of $\hat{\cal H}_{i}$. Through the corresponding Bloch-like equations, its time-evolution is ruled by $\partial_{\tau}{\bf C}_i(\tau)={\bf M}_i\,{\bf C}_i(\tau)$ with ${\bf M}_i$ a square matrix of coeffcients and ${\bf C}(0)_{i}$ easily determined via $\varrho^{ss}_{i}$.
By arranging each ${\bf C}_{i}(\tau)$ so that $(C_{i}(\tau))_{1}=\langle\hat{\sigma}_{+,1}(\tau)\hat{\sigma}_{-,1}(0)\rangle_{ss}$, we arrive at the elegant expression $S(\nu)=\left(\Re\left[(i\nu\openone-{\bf M})^{-1}{\bf C}(0)\right]\right)_{1}$. Analytic solutions can be gathered and turn out to be identical to what is found by brute force solution of the dynamics of ${\bf C}_i(\tau)$'s and Cramer's rule~\cite{wallsmilburn,barnett}. It immediately appears that, regardless of the interaction model being adopted, {\it a discrimination of the parameters characterizing the bath ${\cal B}$ affecting qubit $2$ is possible}. Indeed, let us examine some particularly well-visible situations. 
\begin{figure}[t]
{\bf (a)}\hskip4cm{\bf (b)}
\includegraphics[width=0.2\textwidth]{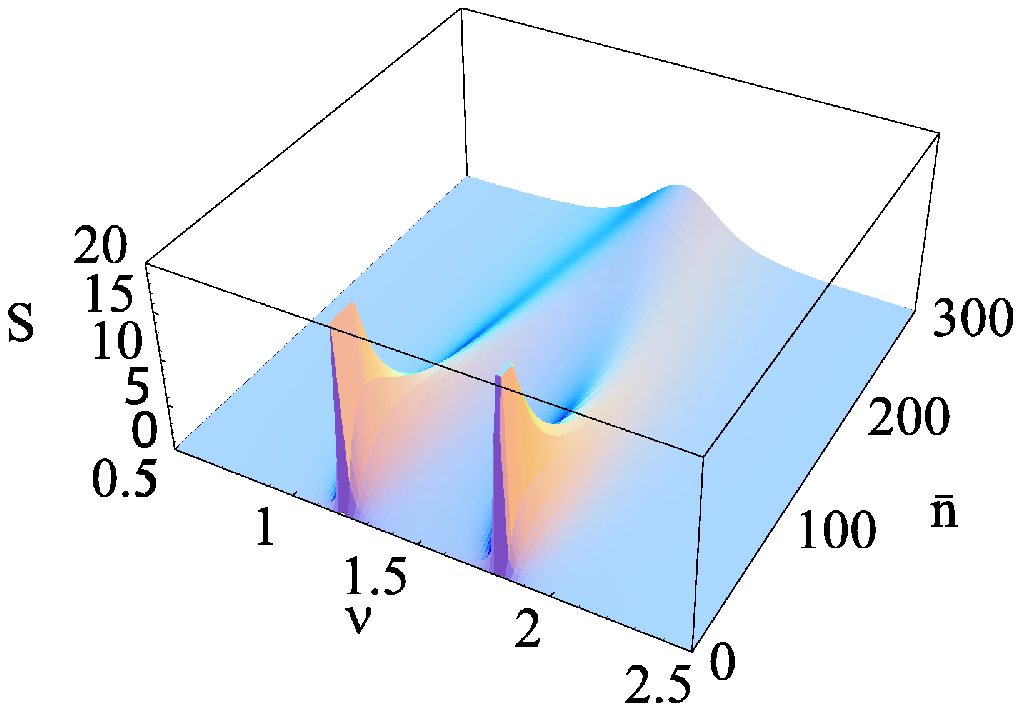}~~\includegraphics[width=0.2\textwidth]{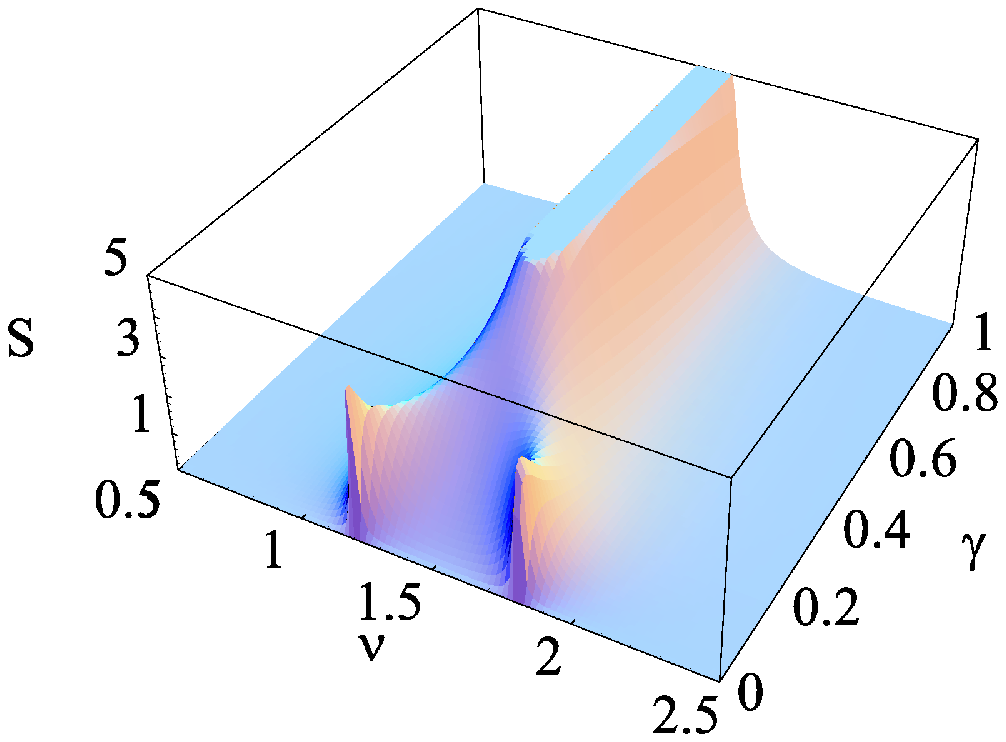}
\caption{{\bf (a)}: $S(\nu)$ against frequency $\nu$ and bath temperature $\nbar$ for the $xy$ model and $J=0.3,\,\delta=0.1,\,\bar\omega=0,\,\gamma=\Gamma=10^{-3},\Omega=3$. {\bf (b)}: Effect of dephasing on the emission spectrum for the same parameters used in panel {\bf (a)} and $\nbar=10^{-2}$.}
\label{fig:fig1}
\end{figure}
Fig.~\ref{fig:fig1} {\bf (a)} addresses the $xy$ model, showing the case of two frequency-degenerate qubit in the presence of a weak bath affecting qubit $2$. Both dissipation and dephasing are considered. Two Lorentzian peaks appear at small temperatures of ${\cal B}$, with a dip in correspondence of the common frequency $\omega_{1}=\omega_2$. This is a clear effect of quantum interference, still possible in such a quasi-unitary situation. As soon as the thermal nature of ${\cal B}$ increases, the incoherent dynamics reduces the dip and merges the Lorentzian peaks into a single thermal peak. The amplitude of the two Lorentzian peaks depends on $\Gamma$ while the dephasing rate $\gamma$ has strong effects on the peak-merging process. For a very small temperature, an increase in $\gamma$ makes the interference effect quickly disappear (see Fig.~\ref{fig:fig1} {\bf (b)}) due to the cancellation of off-diagonal elements in $\varrho^{ss}_{xy}$. On the other hand, larger dissipation implies that the amplitude of the peaks fades more rapidly (an increase in $\Gamma$ from $0.001$ to $0.01$ needs a reduction of the $\nbar$ range from $300$ to $30$). That is, the engineered interaction enables the discrimination of different environmental conditions to which qubit $2$ is exposed. We can infer whether ${\cal B}$ is dissipative (dephasing) only: for instance, $S(1.5)$ at $\Gamma=0.15,\nbar=10$ and $\gamma=10^{-3}$ is roughly six times smaller than the value corresponding to the case of $\Gamma\leftrightarrow{\gamma}$. We can have even more insight and discern if any asymmetry is present in the free dynamics of the qubits. Indeed, for $\bar\omega\neq{0}$, the spectrum becomes asymmetric, with one of the peaks more pronounced and fading more slowly than the other one. Analogous features are found in Figs.~\ref{fig:fig5}, where the $cbf$ model is studied. 

In the case of Fig.~\ref{fig:fig1} {\bf (a)}, $N^{ss}_{xy}$ disappears for $\nbar\ge{0.03}$ while for Figs.~\ref{fig:fig5}, strikingly, we know that $\varrho^{ss}_{cbf}$ is never entangled. Nevertheless, the process of environment discrimination is effective. To fix the ideas, let us concentrate on the $xy$ case: the $\varrho^{ss}_{xy}$ associated with Fig.~\ref{fig:fig1} {\bf (a)} is not diagonal but keeps (small) coherences. Although at most classical correlations are shared by the qubits (as witnessed by the fact that the mutual information in $\varrho^{ss}_{xy}$, accounting for the ``total'' correlations between the qubits, is non-zero), as long as the interaction is present (and not overcome by decoherence killing even the classical correlations in the system), the environment can be characterized. Indeed, for the $xy$ model with $J_{x,y}=0$ ($cbf$ model with $\vartheta=0$), no inferrence from the observation of $1$ is possible (the emission spectrum is flat), even in the presence of preconstituted entanglement. However, any small deviation from this condition brings the probing mechanism back to effectiveness, regardless of the absence of entanglement. Only the {interaction}
rules the possibility of probing the environment.

\begin{figure}[b]
{\bf (a)}\hskip4cm{\bf (b)}
{\includegraphics[width=0.2\textwidth]{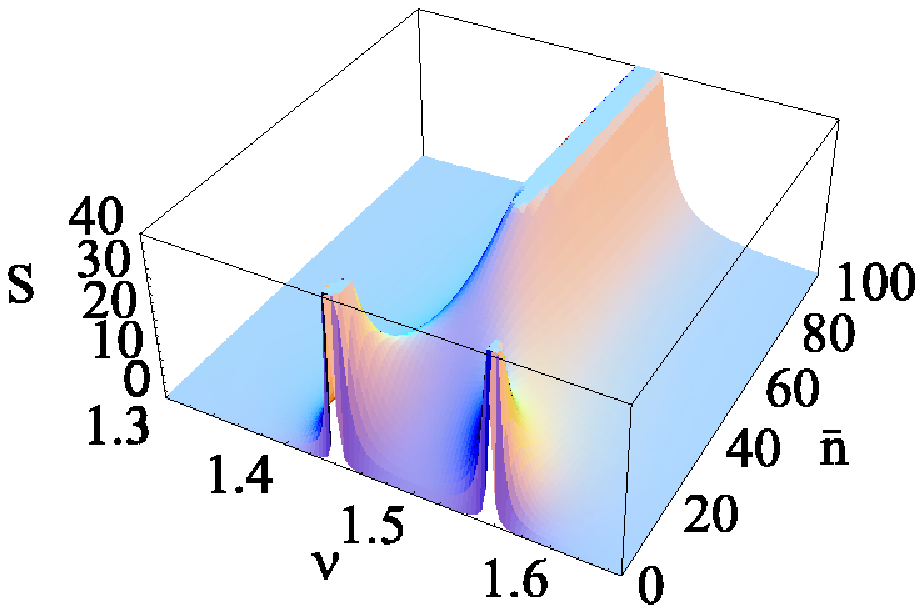}}~~{\includegraphics[width=0.2\textwidth]{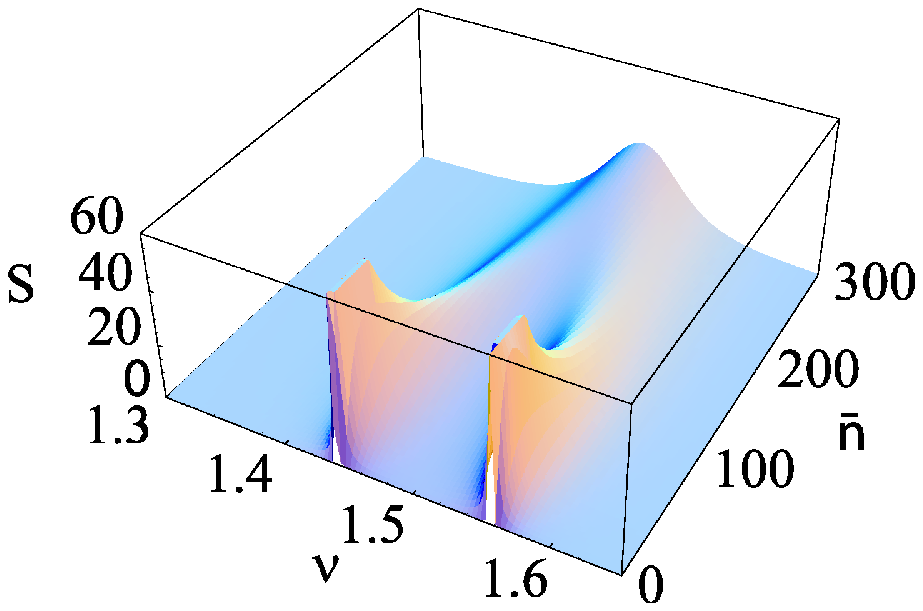}}
\caption{{\bf (a)}: $S(\nu)$ vs. $\nu$ and $\nbar$ for the $cbf$ model with $J=0.3,\,\bar\omega=0,\Omega=3,\gamma=\Gamma=0.001$. {\bf (b)}: Same plot, for $\gamma=\Gamma=10^{-4}$. The behavior against $\gamma$ is similar to Fig.~\ref{fig:fig1} {\bf (b)}.}
\label{fig:fig5}
\end{figure} 

{\it CV case}.- The results described so far can be extended to the CV scenario without any changes to our conclusions. Rather than slavishly reproduce the approach followed in the qubit case, here we show that entanglement does not imply the ability to infer the properties of an environment affecting an inaccessible part of a bipartite CV system. Let us consider a two-mode squeezed state of squeezing factor $r$ (the modes are labelled $1$ and $2$). While mode $1$ undergoes a unitary free evolution, mode $2$ interacts with a bath that, without affecting the generality of our discussions, is taken as dissipative and characterized by the (unknown) parameters $(\Gamma,\nbar)$. No interaction is assumed to connect the two modes. This restricts the study to a dissipative dynamics best described through the Markovian Fokker-Planck equation (FPE) for the Wigner function of the two-mode state~\cite{barnett}.
The solution of the FPE is the convolution of the initial Wigner function of a two-mode squeezed state~\cite{barnett} and that of a thermal state 
modelling the environment~\cite{myungFP}. In details
$W_{}(\alpha,\beta,\tau)
={\cal N}{\exp}[{-\frac{E_\alpha}{D}|\alpha|^2-\frac{E_\beta}{D}|\beta|^2+\frac{4F}{D}(\alpha\beta+\alpha^*\beta^*)}]$,
where ${\cal N}$ is a normalization factor, $T=\sqrt{1-R^2}={\exp}[{-\Gamma{\tau}/2}]$, $D=2T^2+2R^2N\cosh2r$. Here, $E_{\alpha}=T^2\cosh{2}r+NR^2,\,E_{\beta}=\cosh{2}r$ and $F=T\sinh{2}r$ 
while $\beta$ ($\alpha$) is the complex quadrature variable of mode $2$ ($1$). By integrating $W_{}(\alpha,\beta,\tau)$ over $\beta$, we obtain the thermal Wigner function $W_{1}(\alpha,\tau)=({2}/{\pi{E}_{\beta}})\exp[-\frac{2|\alpha|^2}{E_\beta}]$ which carries no information about the environment affecting mode $2$, confirming that entanglement is not sufficient for environmental probing.  

In order to address the case of two interacting CV's, we consider a non-trivial case of experimental interest. Let us reason {\it a posteriori} by considering the radiation-pressure Hamiltonian $\hat{\cal H}_{rp}=\chi\hat{a}^\dag_{1}\hat{a}_{1}(\hat{a}^\dag_2+\hat{a}_2)$ responsible for the coupling between a field mode and the vibrating mirror of an optical cavity (modelled as a harmonic oscillator). Here, $\hat{a}_1$ ($\hat{a}_2$) is the annihilation operator of the field (mirror) and $\chi$ is the coupling constant~\cite{micro,mauronjp,sougato}. The time-propagator ${\cal U}_{rp}={\rm e}^{-i\chi{t}\hat{n}_a(\hat{b}^\dag+\hat{b})}$ is a displacement operator~\cite{barnett} whose complex amplitude depends on the intensity of the field. If the Hilbert space of mode $b$ is restricted to the subspace with at most one excitation, ${\cal U}_{rp}$ is isomorphic to ${\rm e}^{-i\hat{\cal H}_{cbf}t}$. $\hat{\cal H}_{rp}$ is thus the CV extension we are looking for. The Bloch-like equations we have discussed in the qubit case can be cast here in terms of appropriate Langevin equations for the quadrature operators of modes $1$ and $2$. Mode $1$ is affected by white noise entering the cavity and mode $2$ experiences quantum Brownian motion due to the contact of the mirror's vibrational mode with background phonons~\cite{mauronjp}. An analytical formalism for the treatment of this situation has been developed~\cite{mauronjp} and used to show that, by measuring the spectrum of the field, complete information regarding the temperature and damping rate of the mirror can be gathered~\cite{micro,mauronjp}. For our purposes, it is important to stress that in the working conditions of the experiment in~\cite{micro}, no entanglement is set between modes $1$ and $2$, therefore putting the framework of Ref.~\cite{mauronjp} within the context of environmental-probing without entanglement. This case is doubly relevant as it describes a detecting stage (the field) also affected by environmental effects (the white noise entering the cavity), therefore demonstrating the validity of our approach for non-unitary dynamics of the ``detector''. 

{\it Remarks}.- In the characterization of the environment affecting a partially-accessible register, entanglement shared with a detecting device is not crucial. Even classical correlations, set through a time-independent interaction, help in identifying the nature of open-system dynamics and we have shown how an experimentally achievable figure of merit can be used. This result, achieved by studying the stationary state of the register-detecting stage system, is independent of the dimension of the Hilbert space and holds in a minimum-control scenario. We used our experimentally-oriented study to reinterpret the recent achievement of control in some mesoscopic systems~\cite{micro} as genuine interaction-enabled environmental probing.

{\it Acknowledgments}.- MP thanks C. Di Franco for discussions. We acknowledge support from the UK EPSRC, the QIPIRC and The Leverhulme Trust (ECF/40157).


\begin{thebibliography}{99}

\bibitem{schoelkopf} A. Wallraff {\it at al.}, Nature (London) {\bf 431}, 162 (2004); D.I. Schuster {\it et al.}, {\it ibidem} {\bf 445}, 515 (2007).

\bibitem{kouwenhoven} F.H.L. Koppens, {\it et al.}, Nature (London) {\bf 442}, 766 (2006).

\bibitem{schwab} M.D. LaHaye, {\it et al.}, Science {\bf 304}, 74 (2004); A. Naik, {\it et al.}, Nature (London) {\bf 443}, 193 (2006).

\bibitem{micro} S. Gigan {\it et al.} Nature (London) {\bf 444}, 67 (2006).

\bibitem{molecule} L. Venkataraman, {\it et al.}, Nature (London) {\bf 442}, 905 (2006); J. Park, {\it et al.}, {\it ibidem} {\bf 417}, 722 (2002).

\bibitem{greiner} M. Greiner, {\it et al.}, Nature (London) {\bf 415}, 39 (2002).


\bibitem{rau} I. Rau, G. Johansson, and A. Shnirman, \prb {\bf 70}, 054521 (2004).

\bibitem{myungnonmark} See J. Lee, {\it et al.} \pra {\bf 70}, 024301 (2004); D. Ahn, J. Lee, M.S. Kim, and S.W. Hwang, \pra {\bf 66}, 012302 (2002). 

\bibitem{imamoglu} A. Imamoglu, {\it et al.}, \prl {\bf 83}, 4204 (1999); M. Paternostro, {\it et al.}, \pra {\bf 71}, 042311 (2005).

\bibitem{wilson-rae}  M. Paternostro, G. Falci, M. Kim, and G.M. Palma, Phys. Rev. B {\bf 69}, 214502 (2004); I. Wilson-Rae and A. Imamoglu, Phys. Rev. B {\bf 65}, 235311 (2002).

\bibitem{wallsmilburn} D.F. Walls and G.J. Milburn, {\it Quantum Optics} (Springer, Heidelberg, 1994).

\bibitem{myungDNS} J. Lee {\it et al.}, \jmo {\bf 47}, 2151 (2000).





\bibitem{mauronjp} M. Paternostro, {\it et al.}, New J. Phys. {\bf 8}, 107 (2006).

\bibitem{barnett} S.M. Barnett and P.M. Radmore, {\sl Methods in Theoretical Quantum Optics} (Oxford, New York, 1997).

\bibitem{myungFP} M.S. Kim and N. Imoto, Phys. Rev. A. {\bf 52}, 2401 (1995); H. Jeong, J. Lee and M.S. Kim, Phys. Rev. A {\bf 61}, 052101 (2000).

\bibitem{sougato} S. Bose, K. Jacobs, and P.L. Knight, \pra {\bf 56}, 4175 (1997).

\end{thebibliography}
\end{document}